\documentclass[12pt,preprint]{aastex}
\usepackage{graphicx}
\usepackage{epstopdf}
\shorttitle{Fluorescent Extraterrestrial Particle Collector}
\shortauthors{Dominguez et al.}

\begin{document}


\title{A Fluorescent Aerogel for Capture and Identification of Interplanetary and Interstellar Dust}


\author{Gerardo Dom\'{\i}nguez\altaffilmark{1} and Andrew J. Westphal} %\altaffilmark{2}}

\affil{Space Sciences Laboratory, University of California,
    Berkeley, CA 94720}

\author{ Mark L.F. Phillips}
\affil{Pleasanton Ridge Research Corporation, Hayward, CA 94542}

\and

\author{ Steven M. Jones}
\affil{Jet Propulsion Laboratory, California Institute of Technology, Pasadena, CA 91109}


\altaffiltext{1}{Department of Physics, University of California, Berkeley}

\begin{abstract}
Contemporary interstellar dust has never been analyzed in the laboratory, despite its obvious astronomical importance and
its potential as a probe of stellar nucleosynthesis and galactic chemical evolution.   Here we report the discovery of a novel fluorescent aerogel which is capable of capturing hypervelocity dust grains and passively recording their kinetic energies. An array of these ``calorimetric" aerogel collectors in low earth orbit would lead to the capture and identification of large numbers of interstellar dust grains.
\end{abstract}

%\keywords{astrochemistry---instrumentation: detectors, miscellaneous---interplanetary medium---ISM: dust, extinction---meteors---techniques: image processing}

\section{Introduction}
Interstellar dust is an important component of the interstellar medium.
Dust dominates the opacity from the far ultraviolet through the far
infrared and hence controls the spectral appearance of most interstellar
objects. Because of dust shielding against dissociating FUV radiation,
molecules can form in dense clouds which allows cooling to low
temperatures and thus, eventually, gravity to overwhelm pressure support
and the formation of new stars. Small dust grains also dominate the
heating of the interstellar gas through the photoelectric effect and
hence controls the structure of the interstellar medium. Despite some 50+
years of active research, the composition of interstellar dust is still
largely guessed at. In essence, our ignorance reflects the difficulty to
infer dust composition from remote astronomical observations. Here we
propose a novel collection agent which allows the discriminatory
collection of interstellar grains and separation from solar system debris.
This promises to open up a new window on the solid component of the
interstellar medium.

Although it is known that IS dust penetrates into the inner solar system
 \citep{grun,amor}, to date not even a single contemporary IS grain has
been captured and analyzed in the laboratory.  Using sophisticated chemical separation techniques,  certain types of refractory ancient IS particles (so-called ``pre-solar grains'') have been isolated from chondritic meteorites (e.g. \citep{amari1}).   Isotopic abundance patterns within these
individual grains often differ wildly from solar-system values, and point
to the formation of these grains in specific astrophysical environments
such as supernova ejecta and the winds of Asymptotic Giant Branch (AGB)
stars.  But because only the most chemically robust particles (e.g., graphite, SiC, Al$_2$O$_3$) survive the harsh chemical separation, this sample is extremely biased, and it is unlikely that these particles are typical of those found in the interstellar medium.  A sample of IS dust collected by spacecraft in the inner solar system would be less biased, and could lead to the first laboratory characterization of the ``typical'' IS dust particle.
Furthermore, such a sample would allow us to detect isotopic, elemental, and mineralogical differences between dust in the protosolar cloud and dust currently residing in the local ISM, and to probe galactic chemical evolution over the 4.6 Gy since the formation of the solar system.  Pre-solar grains are already proving to be a valuable probe of galactic chemical evolution and stellar nucleosynthesis \citep{amari2,nittler}.

The vast majority ($\sim84$\%) of large ($>1\mu$m) ancient pre-solar grains
appear to be of one type,  so-called ``mainstream'' SiC grains.  These
grains are enriched in $^{22}$Ne,  show s-process signatures in Kr and Xe, and probably originate in the outflows of AGB stars.  Grains from other astrophysical sites have been identified but are  relatively rare (e.g., Type-A,B SiC, tentatively identified with  J-type  carbon stars \citep{amari3}, 3-4\%; Type-X SiC from supernovae \citep{amari1},  1\%;  and alumina from high-metallicity red giants, $<0.5$\%.)  A few pre-solar grains show isotopic patterns that are unique among the thousands that have been  studied so far \citep{nittler}.  If dust in the local ISM shows a similar pattern of diversity, with a dominant common type and relatively rare populations of exotic grains, a large-statistics collection technique will be required to capture, identify and study contemporary IS dust grains from  a wide variety of astrophysical sources.

Aerogels are extremely low-density solids whose superiority  as capturing media for hypervelocity ($v>0.5$km/s)  grains has been well established \citep{barrett,horz2,kitazawa}.  A prominent example is the use of silica aerogel as the collecting medium for cometary and interstellar grains on NASA's Stardust mission \citep{stardust}.  Aerogel collectors have been deployed in low-earth orbit, but severe background from anthropogenic orbital debris has so far prevented the  identification of more than a handful of interplanetary particles \citep{horz2}.  No interstellar particles have been identified so far.  Since they are on hyperbolic orbits, extraterrestrial particles are faster than orbital debris, so could in principle be identified on that basis, but existing aerogels give little information on impact velocity.  With this in mind, we have developed a novel calorimetric aerogel which passively records the kinetic energy of captured hypervelocity particles.

The capture of a hypervelocity dust particle in aerogel produces a shock wave that deforms, heats, and vaporizes the aerogel material in the vicinity of the projectile's trajectory,  resulting in the formation of a permanent track.  The correlation between captured projectile velocity and track characteristics (e.g., track length, track volume, etc.) is poor \citep{kitazawa}.  This behavior is expected theoretically \citep{anderson,westphal-phillips}(G. Dom\'{\i}nguez in preparation). The amount of local heating, however, is nearly linearly proportional to the projectile kinetic energy \citep{anderson}.  If this local heating alters some property of the aerogel in the vicinity of the track, then this property could be used as a calorimeter.  We chose to focus on inducing a fluorescence signal.

\section{Observation of Fluorescence from Capture Events}

We have observed fluorescence resulting from the thermal alteration of aerogels previously in various doped aerogel systems, which fluoresce weakly in their amorphous state and strongly when  baked at high temperatures ($\simeq 1000^{\circ}$C) for extended periods of time ($\sim 1 $hr).  A simple example of such a system is alumina aerogel doped with chromium (III). The amorphous, unheated phase is only very weakly fluorescent under UV illumination (254 nm or 365 nm). Heating the aerogel to 1450$^{\circ}$ C causes it to crystallize to the well-known luminescent phase $\alpha$-Al$_2$O$_3$:Cr, known in Nature as ruby, which glows red ($\lambda_{max}\simeq 700 $nm) under UV illumination. More complex systems include alumina gels co-doped with Gd and Tb. Gd acts as a sensitizer by absorbing UV light at certain wavelengths and nonradiatively transferring energy to Tb, which emits at several wavelengths, principally in the green.

Local heating that results from the capture of hypervelocity projectiles is rapid and confined to small regions in the aerogel.  However, the inducement of a fluorescent state as a result of rapid (t$<200\mu$s), local heating (within $<100\mu$m of the particle track) in an aerogel has previously not been reported.  To test whether local heating  in an aerogel could induce an irreversible phase transformation into a fluorescent phase, the effects of hypervelocity projectile capture were first simulated by exposing samples of Cr-doped and (Gd,Tb)-doped alumina aerogels ($\rho\sim$170 mg/cc) with  a pulsed CO$_2$ laser (300 Hz, 50$\mu$m spot size, pulse width$=$50$\mu$s, power$=$0.25-0.50 W).  The energy per pulse is approximately the energetic equivalent of a glass sphere 10 microns in diameter impacting at 10 km/s.  Some of these aerogels displayed brilliant green fluorescence in the regions of local heating.  This was encouraging evidence that the capture of hypervelocity dust particles could induce a fluorescent phase in alumina aerogels.  These alumina aerogel samples were selected for shots with hypervelocity projectiles (a mix of powdered meteorite and glass beads) at the Advanced Vertical Gun Range at  NASA Ames Research Center.  Two of these samples showed intense green fluorescence in the heated material surrounding the particle tracks, thus establishing that the phase transformation occurs in alumina aerogels.  Quantitative measurements with these shots were precluded because of the large spread in particle sizes and the unknown effect of particle ablation.  These shots were followed more recently, again at Ames, with projectiles consisting of a mixture of monodisperse glass spheres.  This allowed us to do quantitative measurements of the fluorescence yield  as a function of particle size and velocity.

%\noindent{\bf }
%\begin{figure}
%\begin{center}
%\resizebox{!}{5cm}{\includegraphics*{/users/domi/aerogels/Images/Axiophot/1-29ubshot4_10x_8_noback.eps}}
%\caption[ames-dec]{Images of aerogel surfaces after exposure to hypervelocity projectiles ($v=4.71$ km/s) of various sizes and compositions.  Fluorescence was excited at 365 nm using a standard fluorescence microscope at 10x.  These tests demonstrated that hypervelocity dust grain capture can induce an irreversible phase transformation in rare-earth doped aerogels. } \label{ames-dec}
%\end{center}
%\end{figure}

\section{Analysis of Fluorescence Observations}

\begin{figure}
\plotone{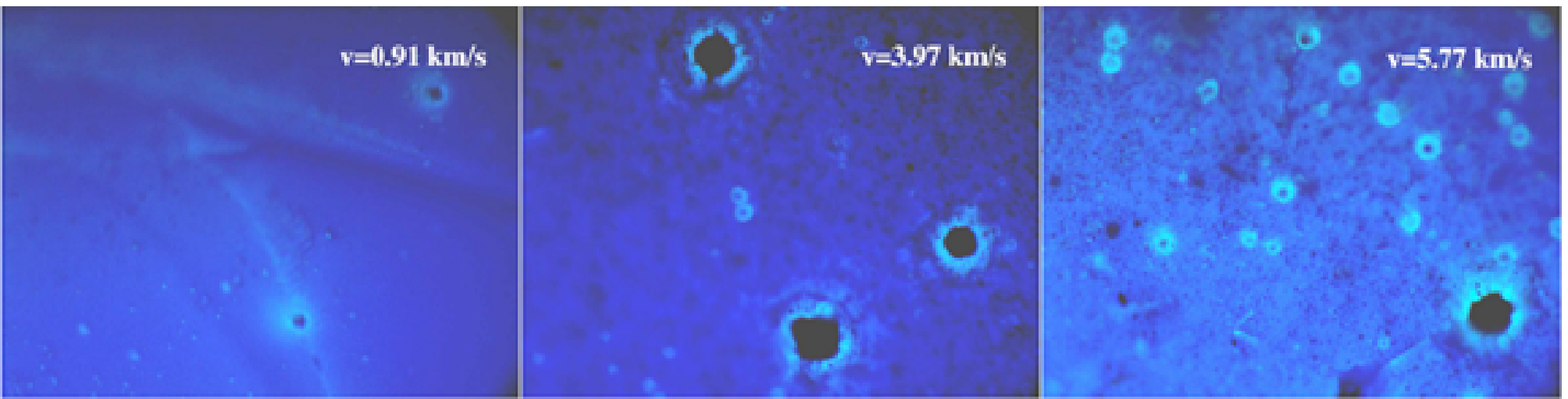}
\caption[ames-june]{Fluorescence images of the surfaces of a Gd,Tb-doped alumina aerogel shot with monodisperse glass (diameters$=$2, 5, and 20 $\mu$m) beads at various velocities at NASA Ames.  Images were taken with a Zeiss Axiophot fluorescence microscope with an attached Optronics DEI-750 3 chip analog camera.  Image contrast was enhanced to improve visibility of fluorescence.} \label{tileofshotswtext}
\end{figure}

\begin{figure}
\plotone{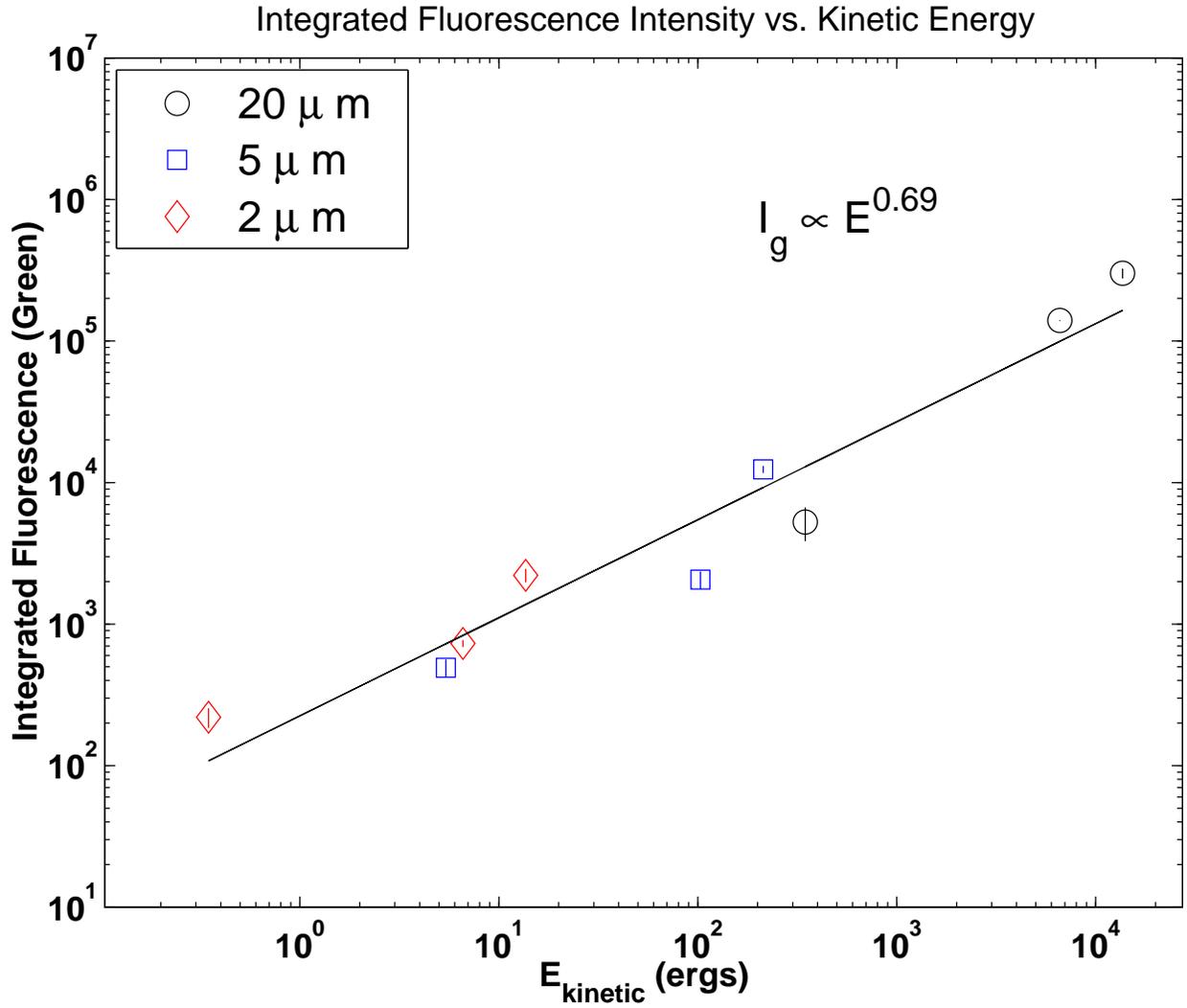}
\caption[calorimetry]{  $I^{net}_{g}$  vs. Kinetic Energy of Hypervelocity Projectiles.  The error bars are statistical only.} \label{IvsEkin}
\end{figure}

We measured the fluorescence yield  using a standard fluorescence microscope with a cooled color CCD video camera.  The fluorescence was excited at 365 nm  using a standard bandpass filter cube at the excitation side and imaged using a long pass filter ($\lambda\ge 395 $nm).  The samples were imaged within two hours of each other to minimize the effects of UV lamp intensity variations.  High resolution images of the aerogel surface where tracks entered were taken and the background fluorescence (weak and mostly blue) was subtracted as follows.  A local blank region of aerogel was sampled, and the average ratio of green to blue, $\overline{f_{gb}}$ was determined; for each pixel we defined the net fluorescence in the green as:
\begin{equation}
		I^{net}_{green}=I_{green}-\overline{f_{gb}}I_{blue}
\end{equation}
where $I_{blue}$ is the blue pixel value.  We chose this background subtraction method because a linear increase in both  the green and blue channels would be expected, even in the absence of a phase transformation, 
due to the increased density of aerogel in the vicinity of the track mouth.  We define the fluorescence yield as the sum of $I^{net}_{green}$ for $I^{net}_{green}>2.5\sigma$ above the pixel noise in the region surrounding the track mouth.   The yield increases dramatically with increasing velocity within each particle population (Fig. \ref{tileofshotswtext}). In Fig. \ref{IvsEkin}, we show the fluorescence yield as a function of kinetic energy. Over the range from 2$\mu$m to 20$\mu$m (three orders of magnitude in mass), the fluorescence yield appears to be consistent with being a single-valued function of the particle kinetic energy, $I_g\propto E_{k}^{0.69}$.  We found that the exponent is insensitive to the choice of fluorescence signal-to-noise threshold.

A reasonable model for the energetics of grain capture can be used to explain, at least qualitatively, the calorimetric aspects of the aerogel.  In this model, we treat the aerogel as a fluid.  In the limit of large Reynolds number, the energy deposited per unit path length by a grain of radius $r$, density $\rho_g$, and  kinetic energy $E$ is 
\begin{equation} {dE\over dx} \sim {3\over2} {1\over r} {\rho_a\over \rho_g} E = {E\over \lambda},\end{equation}
where $\rho_a$ is the aerogel density, and 
\begin{equation}
\lambda = {2\over3} r {\rho_g\over\rho_a}.\end{equation}
This stopping length scale agrees to within 10\% of the value obtained following the more detailed treatment by Anderson and Ahrens \citep{anderson}.  The range of the particle in its supersonic slowing phase $R$ is 
\begin{equation} R_{\rm super} \sim 2\lambda \ln\left({v\over v_{\rm sonic}}\right),
\end{equation}
 where $v_{\rm sonic}$ is the speed of sound in the aerogel.  The logarithmic dependence of the supersonic range on velocity is consistent with the weak dependence observed experimentally \citep{kitazawa}.  If some fraction of the energy loss contributes to the local heating of the aerogel, we should expect the amount of aerogel crystallized to increase as the projectile kinetic energy increases.  Assuming that the luminescence we observe is dominated by one fluorescent phase, the mass per unit track length that is converted into this fluorescent phase is expected to be
\begin{equation}\label{dmdx}
	\frac{dm_{fl}}{dx}\propto v^{2}r^{2}.
\end{equation}

The dependence of $I_g$ on the amount of crystallized aerogel is not necessarily straightforward, as it depends on the optical properties (ultraviolet and visible) of the aerogel as well as the track length.   For events with large track lengths, such as those due to 20 $\mu$m diameter grains, the fluorescence yield may be dominated by the fluorescence at or near the surface of the aerogel.  If so, then $I_{g}$ should be proportional to the amount of aerogel crystallized near the track entrance.  For constant density, therefore
\begin{equation}
	I_{g}\propto m^{\frac{2}{3}}v_{0}^{2}. 
\end{equation}
where $v_{0}$ is the initial impact velocity.  Conversely, for shallow events, the fluorescence yield is expected to be more sensitive to the total amount of aerogel crystallized, hence we would expect $I_g\propto E_{0}$.  Interestingly, a fit of  $I_g$ versus $m^{\frac{2}{3}}v_{0}^{2}$ appears to be a better fit to the data than $I_g$ versus $E$.      Regardless of which functional form ($E_0$ or $E_{0}m^{-\frac{1}{3}}$) turns out to be more accurate when additional studies are done, the main implication of the results reported here is the same.  Together with an independent measure of the mass, for example using {\em in situ} optical imaging or x-ray fluorescence, the velocity of the embedded projectile can be determined \citep{flynn1,flynn2}.

The expectation that the amount of crystallized aerogel scale with the amount of heat deposited is not obvious.  The dependence of emission intensity on grain kinetic energy is complicated by the fact that several phase transitions can occur as alumina aerogels are heated (amorphous $\rightarrow$ $\gamma$-Al$_2$O$_3$ $\rightarrow$ $\theta$-Al$_2$O$_3$ $\rightarrow$ $\alpha$-Al$_2$O$_3$ (corundum)) \citep{mizushima-hori}.  Fluorescence efficiencies of dopant ions typically depend strongly on their local crystal field, and thus it is likely that even if higher grain kinetic energy does not crystallize a larger mass of aerogel, the higher temperatures produced within the track will yield phases of different luminescence.  For example, the heating of Al$_2$O$_3$:Gd,Tb doped aerogels to even moderate temperatures ($\sim$ 1100$^{\circ}$ C) can precipitate phases such as the perovskite phase GdAlO$_3$:Tb and the meta-stable garnet phase Gd$_3$Al$_5$O$_{12}$:Tb, which yield brilliant green luminescence under UV illumination.

\begin{figure}
%\plotone{analysistrack7.eps}
\plotone{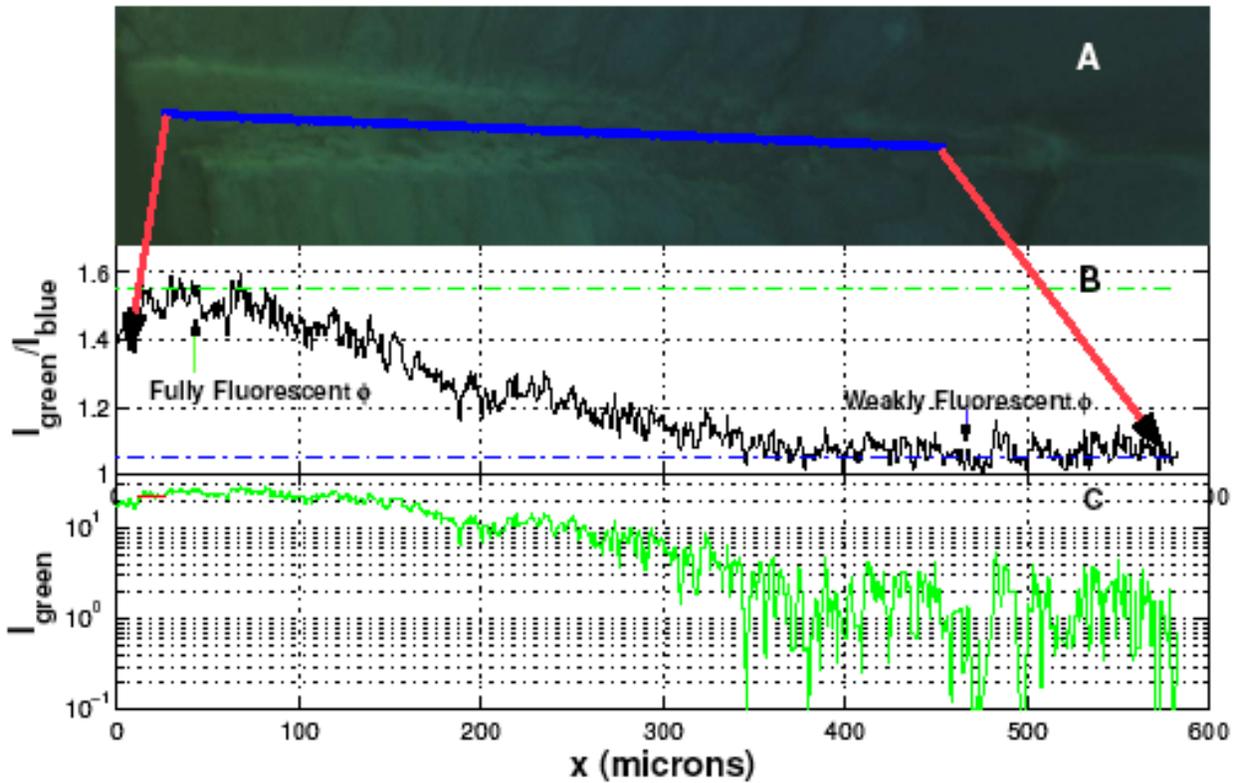}
\caption[1-29]{\textbf{A}.  Fluorescence image of a cleaved track produced by a particle $\simeq 30\mu$m in diameter and initial velocity equal to 4.72 km/s. \textbf{B}. The ratio of green to blue, indicating the degree to which the sampled track region has been transformed into the fluorescent phase. \textbf{C}. Fluorescence signal in the green channel sampled along the track.  Notice that the fluorescence drops off significantly at the end of the track.  Image contrast was not enhanced.} \label{cleavedtrack}
\end{figure}

The kinetics of crystallization in doped alumina aerogels are not known but a more detailed understanding should be helpful in maximizing their usefulness.  The density of the aerogels used in this study are higher than optimalÊ for capture of small hypervelocity particles.ÊÊ Developing larger samples withÊ lower density is a major focus of our current work. This will allow us to characterize the optical properties of our calorimetric aerogel, which will in turn help us understand theÊ effectsÊ that oblique impacts and particle fragmentation may have on integrated fluorescence.Ê Studies of the crystallographic phase and fluorescence dependence on temperature could be used as an {\em in situ} temperature gauge along tracks and could improve our understanding of the kinetics of grain capture in aerogels. Detailed, systematic studies of \emph{fluorescence along a track} may provide information on the instantaneous energy loss that a captured dust particle experiences (see Fig. \ref{cleavedtrack}).Ê

\section{Discussion}

\begin{figure}
\plotone{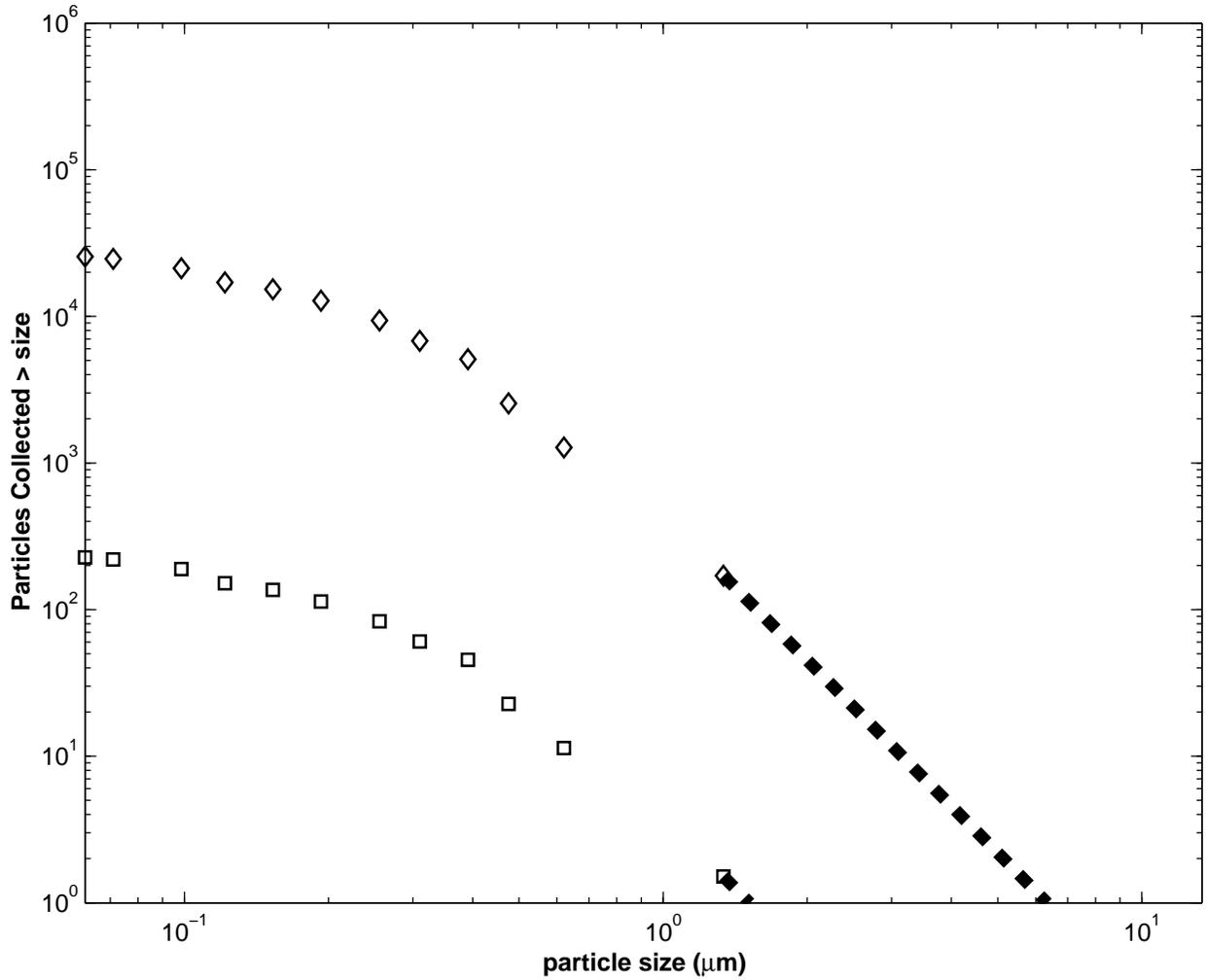}
\caption[isflux]{Cummulative number of interstellar dust grains vs. size  expected to be collector in low-earth orbit ($\Diamond$) compared to the Stardust mission ($\Box$).  The collecting area of the low-earth orbit collector is assumed to be $=$3 m$^{2}$, exposed for 3 years (30 \% duty cycle).  The extrapolated portion of the calculation assumes an IS flux that falls off as $m^{-1.1}$. }
 \label{xpcvsstardust}
\end{figure}

The Stardust spacecraft, whose primary mission is to return samples of
cometary dust
to Earth for laboratory study, has exposed aerogel collectors to the
interstellar dust stream
during two periods of its cruise phase.    The Stardust collectors
will be returned in 2006.  Models of the IS dust flux in the inner solar system indicate that
the Stardust collectors will capture $\sim10$ 1-$\mu$m particles, and
perhaps one 2-$\mu$m particle.
An array of calorimetric aerogel, with collecting area of 3 m$^2$
deployed in low earth orbit for two years,  would have enough collecting
power to collect several hundred 1-$\mu$m  IS particles \citep{landgraf}.
A collector deployed on the wake side of a spacecraft in low-earth orbit could collect
IS dust at moderate velocites ($< 10$ km/s) during periods of the year when the earth's motion
is most parallel to that of the IS dust stream \citep{grun}.
Furthermore, the largest particle expected to be captured by such an array would be  $\sim$30 times more massive than the largest particle expected to be collected by Stardust \citep{landgraf} (see Figure \ref{xpcvsstardust}). These particles would be large enough to apply multiple chemical,
mineralogical, and isotopic analysis techniques to each particle \citep{smallisbeautiful}.

The main point of this paper is that we have discovered a method of distinguishing between copious anthropogenic debris and relatively rare extraterrestrial particles captured in a collector in low earth orbit.ÊÊ Although we have chosen to emphasize the importance of capturing large numbers of contemporary interstellar grains for the first time, it is inevitable that interplanetary dust particles would also be collected.Ê Both of these populations would be of scientific interest, and separating these two population is a complex problemÊ and beyond the scope of this paper.ÊÊ A large-statistics collection of interplanetary dust collected in space would be a valuable resource for the meteoritics and planetary science community.Ê So-called Interplanetary Dust Particles (IDPs) have been collected in the stratosphere for many years \citep{brownlee02}.  Micrometeorites have also been collected terrestrially in Antarctica --- these are the so-called Antarctic Micrometeorites (AMMs) \citep{engrand}.  Each of these collection techniques has its own biases. Both are biased towards particles that can survive atmospheric entry.  The effects of atmospheric contamination are poorly understood \citep{flynn95}.  Terrestrially-collected micrometeorites are selected towards particles that survive weathering  and that are readily recognized as extraterrestrial.

A few chondritic particles have been extracted from ordinary silica aerogel collectors flown in space and analyzed \citep{horz2}.  These chondritic grains were selected from a large background of anthropogenic particles.  The relationship between IDPs, AMMs, micrometeorites and ordinary chondrites is not clear \citep{brownlee02,flynn02}, but it appears that AMMs constitute a different population than IDPs, and may have a different  origin.  A single collector that is large enough to capture, in space,  several 100 $\mu$m particles --- characteristic of AMMs --- along with IDPs could clarify the relationship between them.  The origin of AMMs in particular is important since they constitute the greatest contemporary mass input to the earth \citep{gounelle}, and could have contributed a significant amount  of water and organics to the early earth.

\acknowledgments

This work was supported by NASA Grant NAG5-10411.  We thank David King, Christopher J. Snead, Peter Schultz, and the crew of the AVGR for helping make these tests possible.   We thank John Bradley and Xander Tielens for useful discussions.  G. Dominguez would like to thank the NPSC Graduate Fellowship Program for their support.
\\
 
Correspondence and requests for materials should be addressed to G.D.

\clearpage

\end{document}